\documentclass[conference,10pt,a4paper]{IEEEtran}

\usepackage{graphicx}
\usepackage{multirow}
\usepackage[none]{hyphenat}
\usepackage{float}
\usepackage{amssymb}
\usepackage{bbm}

\usepackage[style=ieee,backend=bibtex,sorting=none]{biblatex}
\usepackage{tikz}
\usepackage{pgfplots}
\usetikzlibrary{calc}
\DeclareUnicodeCharacter{2212}{−}
\pgfplotsset{compat=newest}
\usepackage{amsmath}
\usetikzlibrary{patterns,shapes.arrows}
\usetikzlibrary{shapes}
\usetikzlibrary{positioning}
\usetikzlibrary{arrows}
\usetikzlibrary{arrows.meta}
\usetikzlibrary{backgrounds}
\usetikzlibrary{arrows}
\usetikzlibrary{shapes,shapes.geometric,shapes.misc}
\usetikzlibrary{math}
\definecolor{yellow}{HTML}{F4AC4C}
\definecolor{blue}{HTML}{34baeb}
\definecolor{green}{HTML}{79B473}
\usetikzlibrary{matrix}

\usepackage{amsfonts}
\usepackage{cleveref}
\usepackage{siunitx}
\usepackage[acronyms]{glossaries}
\newacronym{ofdm}{OFDM}{orthogonal frequency-division multiplexing}
\newacronym{tx}{TX}{transmitter}
\newacronym{rx}{RX}{receiver}
\newacronym{isac}{ISAC}{integrated sensing and communications}
\newacronym{los}{LoS}{line-of-sight}
\newacronym[plural=FFTs]{fft}{FFT}{fast Fourier transformation}
\newacronym[plural=IFFs]{ifft}{IFFT}{inverse fast Fourier transformation}
\newacronym[plural=DFTs]{dft}{DFT}{discrete Fourier transformation}
\newacronym[plural=FTs]{ft}{FT}{Fourier transformation}
\newacronym[plural=SNRs]{snr}{SNR}{signal-to-noise ratio}
\newacronym[]{ml}{ML}{maximum likelihood}
\newacronym[]{crb}{CRB}{Cramer-Rao bound}
\newacronym[plural=CNNs]{cnn}{CNN}{convolutional neural network}
\newacronym[]{cfar}{CFAR}{constant false alarm rate}
\newacronym[]{cut}{CUT}{cell under test}
\newacronym[]{oscfar}{OS-CFAR}{ordered statistic constant false alarm rate}
\newacronym[plural=CPIs]{cpi}{CPI}{coherent processing interval}
\newacronym[plural=RMSEs]{rmse}{RMSE}{root mean squared error}
\newacronym[]{radar}{radar}{radio detection and ranging}
\newacronym{pinn}{PINN}{physics-informed neural network}
\usepackage[figurename=Figure]{caption}
\usepackage{subcaption}
\usepackage[]{tabularx}
\usepackage{booktabs}

\DeclareMathOperator*{\argmin}{argmin}

\usepackage[activate={true,nocompatibility},final,tracking=true,kerning=true,spacing=true,factor=1100,stretch=30,shrink=30]{microtype}

\DeclareFieldFormat[thesis]{url}{[Online]. \mkbibacro{Available:}\space\url{#1}}
\AtEveryBibitem{
    \clearlist{address}
    \clearfield{month}
    \clearfield{editor}
    \clearfield{eprint}
    \clearfield{volume}
    \clearfield{issn}
    \clearlist{location}
    \clearlist{editor}
    \clearfield{pages}
}

\begin{document}
\title{Performance Comparison\\of Joint Delay-Doppler Estimation Algorithms}
\author{%
    \IEEEauthorblockN{
        Lorenz~Mohr\IEEEauthorrefmark{1}, 
        Michael~D\"obereiner\IEEEauthorrefmark{2},
        Steffen~Schieler\IEEEauthorrefmark{1},
        Joerg~Robert\IEEEauthorrefmark{2},\\
        Christian~Schneider\IEEEauthorrefmark{1}\IEEEauthorrefmark{2},
        Sebastian~Semper\IEEEauthorrefmark{1}\IEEEauthorrefmark{2}, and
        Reiner~S.~Thom\"a\IEEEauthorrefmark{1}
    }
    \IEEEauthorblockA{
        \IEEEauthorrefmark{1} Technische Universit\"at Ilmenau, Institute of Information Technology, Ilmenau, Germany\\
        \IEEEauthorrefmark{2} Fraunhofer Institute for Integrated Circuits IIS, Ilmenau, Germany
    }
} 
\maketitle

\begin{abstract}
\sisetup{propagate-math-font = true, reset-math-version = false}
\Gls{isac}, radar, and beamforming require real-time-capable, high-resolution parameter estimation algorithms to determine delay-Doppler values of specular paths within the wireless propagation channel. 
Our contribution is the measurement-based performance comparison of the delay-Doppler estimation between three different algorithms---comprising \gls{ml}, \gls{cnn}, and \gls{cfar} approaches.
We apply these algorithms to channel data comprising two spherical targets with analytically describable delay-Doppler ground truth.
The comparison features the empirical target detection rate, \glspl{rmse} of the delay-Doppler estimates, and a runtime analysis.
While the \gls{cfar}-based algorithm provides the best computational efficiency, the high-resolution capabilities of the \gls{ml} and \gls{cnn} approaches guarantee higher estimation accuracy and target detection rate.
\end{abstract}
\glsresetall
\section{Introduction}
\label{1-introduction}
\Gls{isac}, radar, and beamforming are important research fields for the sixth-generation mobile communication standard.
This technology necessitates a signal waveform that is capable of transmitting data and sensing the environment simultaneously~\cite{Najaf.2025,Wei.2023, Sturm.2011}.
As Braun~\cite{Braun.2010} and Sturm~\cite{Sturm.2009} demonstrated in previous research activities, \gls{ofdm} fulfills this condition.
They showed that such a signal facilitates delay, Doppler, and joint delay-Doppler estimation, while concurrently maintaining robust communication capabilities~\cite{C.SturmM.BraunT.ZwickandW.Wiesbeck., Braun.2014}.
In addition, \gls{isac} requires high-precision and real-time capable delay-Doppler parameter estimation for sensing~\cite{Semper.2024,Liu.2022, PinTan.}.

To perform such a parameter estimation, a multitude of algorithms exist.
On one hand, there are \gls{dft}-based approaches.
These algorithms estimate the propagation parameters by performing variations of a peak search in the delay-Doppler spectrum~\cite{Braun.2014}.
As the \gls{fft} is a computationally efficient implementation of the \gls{dft}, these algorithms are well-suited for real-time applications at the cost of a finite resolution, limited by the bandwidths in time and frequency~\cite{Richards.2022}.

On the other hand, model-based algorithms estimate the target parameters by fitting an appropriate model to the received data. 
One example here is RIMAX, which performs delay-Doppler estimation based on an iterative \gls{ml} procedure~\cite{Richter.2005}.
Due to the analytical signal model, RIMAX is capable of separating closely spaced paths that are below the \gls{dft} resolution.
Another example are \gls{pinn}-based estimators.
In~\cite{Schieler.2024}, Schieler introduces such an algorithm, which is termed DeepEst within our work.
This algorithm utilizes a \gls{cnn} structure in combination with a second-order gradient iteration to determine delay-Doppler values of specular propagation paths.
However, the increased estimation accuracy of both RIMAX and DeepEst comes at the expense of a higher runtime compared to \gls{fft}-based approaches, resulting in diminished real-time capabilities~\cite{Schieler.2024}.

Within our work, we present a comparison of the joint delay-Doppler estimation performance for three algorithms---RIMAX, DeepEst and \gls{oscfar}---in terms of target detection probability, \glspl{rmse} of the delay-Doppler estimates, and algorithm runtime.
We perform this comparison utilizing channel measurements from a metrologically assessable setup in the sub-6-\si{\giga\hertz} frequency band, including two moving spherical targets.
Consequently, analytical ground truth values for the delay-Doppler propagation parameters of both spheres can be derived, enabling a direct comparison of the algorithms.
\section{Measurement Data}
\label{2-data}


\subsection{Measurement Setup}
\label{2-data/setup}

\begin{figure}[b]
    \centering
    \includegraphics[width=0.5\linewidth]{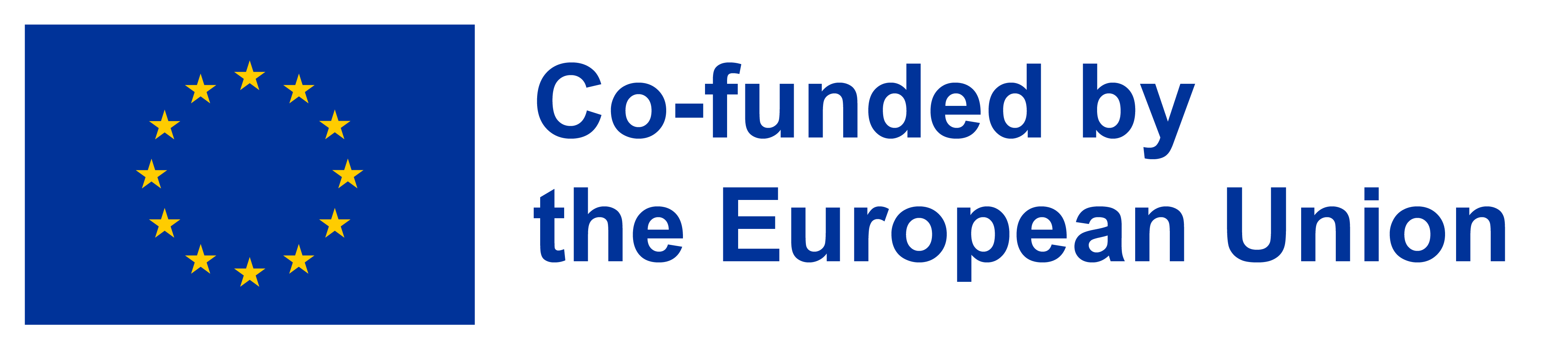}
\end{figure}

\begin{figure}[t]
    \centering
    \definecolor{indianred1967882}{RGB}{196,78,82}

\begin{tikzpicture}[scale=0.8]
    \draw[black, thick] (0,0) circle(3cm);
    \draw[black,thick] (0,-0.1) -- (0,0.1);
    \draw[black,thick] (0.1,0) -- (-0.1,0);
    \draw[black, thick, stealth-stealth] (-3,0) arc(-180:-210:3);

    \draw[black] (0,0) -- (0,-3.3);
    \draw[black] (-0.2, -3.3) -- (0.2, -3.3);
    \draw[black] (-0.2, -3.3) -- (-0.05, -3.7);
    \draw[black] (0.2, -3.3) -- (0.05, -3.7);
    \draw[black] (-0.05, -3.7) -- (-0.05, -3.8);
    \draw[black] (0.05, -3.7) -- (0.05, -3.8);
    \draw[black] (0.05, -3.8) -- (-0.05, -3.8);
    \node at (0.5,-3.5) {TX};

    \draw[black] (0,0) -- (-1.635,-1.635);
    \node at (-1.9,-1.3) {RX};
    \begin{scope}[shift={(0.7,0.7)}, rotate=-45]
        \draw[black] (-0.2, -3.3) -- (0.2, -3.3);
        \draw[black] (-0.2, -3.3) -- (-0.05, -3.7);
        \draw[black] (0.2, -3.3) -- (0.05, -3.7);
        \draw[black] (-0.05, -3.7) -- (-0.05, -3.8);
        \draw[black] (0.05, -3.7) -- (0.05, -3.8);
        \draw[black] (0.05, -3.8) -- (-0.05, -3.8);
    \end{scope}

    \draw[black] (0,-1) arc(-90:-135:1);
    \node at (-0.25,-0.6) {\(\delta\)};

    \draw[black, thick] (-1,1) -- (1,-1);
    \draw[black, -stealth] (-1,1) arc(135:100:1.414); 
    \draw[black, -stealth] (1,-1) arc(-45:-80:1.414);
    \filldraw[indianred1967882] (-1,1) circle(2pt);
    \filldraw[indianred1967882] (1,-1) circle(2pt);

    \draw[black, thin, stealth-] (0.5,-0.5) -- (2.5,3) node[above] {Beam};
    \draw[black, thin, stealth-] (1,-1) + (2pt,2pt) -- (3.5,1.5) node[above] {Sphere};
    \draw[black, thin, stealth-] (3,0) -- (4,0) node[right] {Turntable};
    \draw[black, thin, stealth-] (0.05,0.05) -- (1,3.5) node[above] {Motor};
\end{tikzpicture}
    \caption{\sisetup{detect-all = true}\textit{Bistatic Measurement Setup---Two metallic spheres~(red) are mounted on a rotating beam. The bistatic measurement angle~$\delta$ between \gls{tx} and \gls{rx} creates either backward, forward, or bistatic scattering scenarios. The length of the beam is \qty{3}{\meter} and the distances of \gls{tx} and \gls{rx} to the turntable center are \qty{3.48}{\meter} and \qty{2.86}{\meter}, respectively.}}
    \vspace{-1em}
    \label{fig:3-meas_setup}
\end{figure}

\Cref{fig:3-meas_setup} depicts the measurement setup, which was originally introduced in~\cite{Schwind.2020}.
Two metallic spheres (red circles) are mounted on a metal rod attached to a motor~(+). 
Both \gls{tx} and \gls{rx} include horn antennas.
The bistatic measurement angle~$\delta$ spans between the \gls{tx}, the turntable center, and the \gls{rx}. 
This angle can be freely varied between~\qty{0}{\degree} and~\qty{360}{\degree}.
Under these rotations, three different radar scenarios emerge from the setup---backward scattering~($\delta \approx \qty{0}{\degree}$), forward scattering~($\delta \approx \qty{180}{\degree}$), and bistatic scattering.
In addition, the measurement setup records Cartesian coordinates of \gls{tx}, \gls{rx}, and both sphere center points. 
Under the assumption that the diameters of the spheres are negligible, this setup enables precise calculation of the theoretical propagation parameters---namely delay and Doppler---for the rotating targets.
The equations for these calculations can be found in~\cite{Schwind.2020}. 

\subsection{Signal Model}
\label{2-data/model}

As introduced in~\cite{Schwind.2020}, the measurement setup utilizes a multi-sinus waveform with Newman sequence phases to optimize the crest factor of the sensing signal.
To perform delay-Doppler estimation of the spheres, we accumulate $L \in \mathbb{N}$ subsequent observations of the channel, each containing $K \in \mathbb{N}$ carriers, into a data frame.

By assuming an underlying sampling process of $\Delta f$ in frequency and $\Delta t$ in time, we describe the complex baseband channel transfer function $\mathbf{H} \in \mathbb{C}^{K \times L}$ by means of a parametric model as
\begin{equation}
    \label{eq:2/cir}
    {\mathbf{H}}_{k \ell}=
    \sum_{p=1}^P\gamma_p \cdot \mathrm{e}^{-\mathrm{j} 2\pi k \Delta f \tau_p} \cdot e^{\mathrm{j} 2\pi \ell \Delta t \alpha_p}+{\mathbf{N}}_{k \ell},
\end{equation}
where $\gamma_p \in \mathbb{C}$, $\tau_p$, and $\alpha_p$ denote the $p$-th specular path in terms of weight, propagation delay, and Doppler-shift, respectively.
The array $\mathbf{N} \in \mathbb{C}^{K \times L}$ is assumed to be drawn from an independent and identically distributed zero-mean circularly symmetric Gaussian process and models noisy observations of the channel.
Since ground truth values for the path weights of the spheres are not available, we restrict this work to joint delay-Doppler parameter estimation.
\section{Algorithms}
\label{3-algorithms}

The performance comparison of this work utilizes the estimation results of three algorithms, namely RIMAX~\cite{Richter.2005}, DeepEst~\cite{Schieler.2024}, and a framework based on \gls{oscfar}~\cite{Rohling.1983}. 

\subsection{RIMAX}
\label{3-algorithms/pymax}

The RIMAX algorithm is a \gls{ml}-based estimator~\cite{Richter.2005}.
It estimates the paths within~\eqref{eq:2/cir} through an iterative procedure in which the parameters of all previously detected paths are refined using a gradient-based Gauss-Newton optimization.
Subsequently, RIMAX validates the detected paths against the Cram\'er-Rao bound to assess the validity of the estimated parameters.
If this test is passed, the contribution of the most recently detected path is subtracted from~\eqref{eq:2/cir}.
The algorithm then continues the estimation on the residual data until the Cram\'er-Rao bound criterion is no longer satisfied.

\subsection{DeepEst}
\label{3-algorithms/deepest}

In contrast, DeepEst is a \gls{pinn}-based algorithm that utilizes a convolutional neural network to estimate the parameters of the specular paths within~\eqref{eq:2/cir}~\cite{Schieler.2024}.
The algorithm processes the data through three convolutional layers that perform channel upscaling, data downsampling, and channel downscaling, respectively.
The output of the final convolutional layer is then passed to two linear layers, which estimate the delay, Doppler, and path weight of the propagation paths.
In addition, the algorithm includes a fourth convolutional stage followed by linear layers.
This stage estimates the total number of paths~$P$ and discards unlikely paths, which were identified in the third convolutional block.

\subsection{OS-CFAR}
\label{3-algorithms/cfar}

We employ \gls{oscfar} in combination with selected pre- and post-processing techniques as a \gls{dft}-based baseline algorithm for comparison with RIMAX and DeepEst.
Since the delay-Doppler spectrum exhibits pronounced sidelobes stemming from the limited aperture and process noise, conventional peak-based target detection results in numerous false alarms~\cite{Richards.2022}.
To mitigate this effect, the \gls{oscfar} algorithm estimates the noise level locally using ordered statistics~\cite{Rohling.1983}, thereby reducing the number of false detections.

Although \gls{oscfar} successfully suppresses most false targets caused by side lobes and noise, strong static paths within the measurement setup still pose a problem for this algorithm.
As the bistatic measurement angle approaches~\qty{0}{\degree} or~\qty{180}{\degree}, the \gls{los} component dominates the received signal by several orders of magnitude, effectively shadowing the weaker reflections from the dynamic spheres~\cite{Schwind.2020}.
To address this issue, we preprocess~\eqref{eq:2/cir} using background subtraction to remove static paths and a two-dimensional Hamming window to attenuate sidelobes.
In addition, we apply a two-dimensional quadratic interpolation in postprocessing to increase the accuracy of the estimation.
\section{Estimation Performance Comparison}
\label{4-comparison}

The measurement data for this comparison and our delay-Doppler estimates are publicly available at~\cite{2026_rhino_dataset_Mohr}.
Based on this data, we perform delay-Doppler estimation on $W \in \mathbb{N}$ consecutive frames---each comprising \num{100} symbols---utilizing the hyperparameters of \Cref{tab:4-hyperparameters} for the three algorithms.
For the $w$-th frame, this step yields $\hat{P} \in \mathbb{N}$ multipath component estimates, each including a propagation delay $\hat\tau_p$ and a Doppler-frequency $\hat\alpha_p$.

\begin{table}[ht]
    \caption{Hyperparameters}
    \centering
    \begin{tabularx}{0.5\textwidth}{X l l}
        \toprule
        Hyperparameter & Value & Explanation\\
        \midrule
        \multicolumn{3}{l}{Measurement Data~(\Cref{2-data/model})}\\
        \midrule
        $K$ & \num{1024} & Number of subcarriers \\
        $L$ & \num{100} & Number of symbols \\
        $f_\mathrm{s}$ & \qty{160}{\mega\hertz} & Signal bandwidth \\
        $\Delta t$ & \qty{64}{\micro\second} & Symbol length\\
        \midrule
        \multicolumn{3}{l}{RIMAX~(\Cref{3-algorithms/pymax})}\\
        \midrule
        $P_\text{max}$ & \num{25} & Maximum number of paths\\ 
        $n_{\text{grad}_\text{max}}$ & \num{50} & Maximum number of gradient iterations\\
        \midrule
        \multicolumn{3}{l}{DeepEst~(\Cref{3-algorithms/deepest})}\\
        \midrule
        $n_{\text{grad}}$ & \num{10} & Number of gradient iterations\\
        \midrule
        \multicolumn{3}{l}{OS-CFAR~(\Cref{3-algorithms/cfar})}\\
        \midrule
        $m_\text{ref}$ & \num{15} & Reference window size in delay\\
        $n_\text{ref}$ & \num{7} & Reference window size in Doppler\\
        $r$ & \num{79} & Noise estimate index from (10) of \cite{Rohling.1983}\\
        $\alpha_\text{os}$ & \num{11.39} & Threshold scaling factor from (14) of \cite{Rohling.1983}\\ 
        \bottomrule
    \end{tabularx}
    \label{tab:4-hyperparameters}
    \vspace{-1em}
\end{table}

\subsection{Performance Comparison Metrics}
\label{4-comparison/classification}

Our work utilizes the three metrics---detection probability, delay-Doppler estimate \glspl{rmse}, and runtime---for performance comparison of joint delay-Doppler estimation algorithms.
The calculation of these metrics requires an association between the $\hat{P}$ estimated multipath components and the $T \in \mathbb{N}$ targets, which provide delay-Doppler ground truth information.
To this end, we define the normalized distance between the $p$-th multipath component and the $t$-th target as
\begin{equation}
    d(p,t) = \sqrt{\left(\frac{\hat\tau_p - \tau_t}{\frac{K}{f_\mathrm{s}}}\right)^2 + \left(\frac{\hat\alpha_p - \alpha_t}{\frac{1}{\Delta t}}\right)^2},
\end{equation}
where $\tau_t$ and $\alpha_t$ denote the delay and Doppler ground truth values for the $t$-th target, respectively.
Performing the operation
\begin{equation}
    \hat{p}_t = 
    \begin{cases}
        \argmin_p d(p,t), & \text{if } \min_p d(p,t) \leq \epsilon\\
        \varnothing, & \text{otherwise}
    \end{cases}
    \label{eq:5/association}
\end{equation}
then associates the $\hat{p}_t$-th component with the $t$-th target if there is an estimate within an identification boundary $\epsilon > 0$.

As a result, the empirical target detection probability and delay-Doppler \glspl{rmse} across $W$ data frames, each comprising $T$ targets, write as
\begin{equation}
    P_\mathrm{D} = \frac{1}{WT}\sum_{w=1}^W \sum_{t=1}^T\mathbbm{1}\{\hat{p}_{w,t} \neq \varnothing\} 
\end{equation}
and
\begin{equation}
    \text{RMSE}_\xi = \sqrt{\frac{\sum_{w=1}^W \sum_{t=1}^T \mathbbm{1}\{\hat{p}_{w,t} \neq \varnothing\}(\hat\xi_{\hat{p}_{w,t}} - \xi_{w,t})^2}{\sum_{w=1}^W \sum_{t=1}^T \mathbbm{1}\{\hat{p}_{w,t}\ \neq \varnothing\}}},
    \label{eq:5/rmse}
\end{equation}
respectively, where $\xi$ denotes a placeholder variable for either delay $\tau$ or Doppler $\alpha$, $\mathbbm{1}\{.\}$ represents the indicator function, and the subscripts $w$ and $t$ encode the frame and target indices.
The appearance of the indicator function in the numerator of \eqref{eq:5/rmse} accounts for unsuccessful associations in \eqref{eq:5/association} and excludes these events from the \gls{rmse} calculation.

\Cref{fig:4/detection_prob} illustrates how the identification boundary~$\epsilon$ affects the empirical target detection probability of the two metallic spheres.
Whereas the solid lines represent a boundary of $\epsilon=0.5$, the dotted lines correspond to evaluations performed with $\epsilon = 0.25$.
As the boundary decreases, the empirical target detection probability of each algorithm correspondingly diminishes.

For further evaluation, we utilize a boundary of $\epsilon = 0.5$.
This selection ensures an acceptable target detection probability in bistatic scenarios while simultaneously maintaining a low number of false targets in both forward and backward scattering.

\begin{figure}[h]
    \centering
\begin{tikzpicture}

\definecolor{orchid218139195}{RGB}{218,139,195}
\definecolor{darkslategrey38}{RGB}{38,38,38}
\definecolor{grey14712096}{RGB}{147,120,96}
\definecolor{indianred1967882}{RGB}{196,78,82}
\definecolor{lavender234234242}{RGB}{234,234,242}
\definecolor{lightgrey204}{RGB}{204,204,204}
\definecolor{lightslategrey129114179}{RGB}{129,114,179}
\definecolor{mediumseagreen85168104}{RGB}{85,168,104}
\definecolor{peru22113282}{RGB}{221,132,82}
\definecolor{steelblue76114176}{RGB}{76,114,176}

\definecolor{darkgray176}{RGB}{176,176,176}
\definecolor{lightgray204}{RGB}{204,204,204}

\begin{axis}[
    width=0.85\columnwidth,
    legend cell align={left},
    legend style={fill opacity=0.8, draw opacity=1, text opacity=1, draw=lightgray204},
    axis y line=right,
    axis x line=none,
    xmin=0,
    xmax=360,
    ymin=-20,
    ymax=70,
    ylabel={LoS Path Gain [dB]}
    ]
\addplot [line width=1pt, indianred1967882, opacity=1]
table {%
0 31.7587347520117
10 14.1674654657386
20 -10.6757596918007
30 -8.66287630857032
40 -3.65388938158911
50 -9.06101907424746
60 -9.16360827909492
70 -9.0727258425549
80 3.91837599964266
90 2.2059353649232
100 1.01826086475312
110 -2.58691521574202
120 2.91083601736805
130 25.1031842121823
140 34.2471214911711
150 39.7760061062179
160 42.8502205358495
170 44.7882459411131
180 45.8493791872172
190 44.9024319321634
200 42.5728593679852
210 40.1471196897103
220 36.4370655042584
230 28.8957593882964
240 3.7098294682086
250 -4.47640377238817
260 -2.04198667792176
270 0.681142799936926
280 5.85975754818694
290 1.14758994959562
300 -3.2644252500083
310 -2.76498396873616
320 5.2297766416632
330 -0.497952306962001
340 -2.33120640840996
350 10.3548255902835
360 31.7587347520117
};\label{los}

\addplot[line width=1pt, steelblue76114176, draw=none] coordinates{
    (0,0)
};\label{oscfar_solid}
\addplot[line width=1pt, black, draw=none] coordinates{
    (0,0)
};\label{oscfar_dashed}
\addplot[line width=1pt, mediumseagreen85168104, draw=none] coordinates{
    (0,0)
};\label{pymax_solid}
\addplot[dashed, line width=1pt, black, draw=none] coordinates{
    (0,0)
};\label{pymax_dashed}
\addplot[line width=1pt, peru22113282, draw=none] coordinates{
    (0,0)
};\label{deepest_solid}
\addplot[dashed, line width=1pt, peru22113282, draw=none] coordinates{
    (0,0)
};\label{deepest_dashed}
\end{axis}

\begin{axis}[
width=0.85\columnwidth,
legend cell align={left},
legend style={fill opacity=0.8, draw opacity=1, text opacity=1, draw=lightgray204},
tick align=outside,
tick pos=left,
x grid style={darkgray176},
xlabel={Bistatic Measurement Angle $\delta \;[^\circ]$},
xmajorgrids,
xmin=0.0, xmax=360,
xtick style={color=black},
y grid style={darkgray176},
ylabel={Empirical Target Detection Probability },
ymajorgrids,
ymin=0, ymax=1.3,
ytick style={color=black},
ytick = {0.0, 0.2, 0.4, 0.6, 0.8, 1.0},
yticklabels={0.0, 0.2, 0.4, 0.6, 0.8, 1.0},
axis y line=left
]
\addplot [line width=1pt, steelblue76114176]
table {%
0 0.436708860759494
10 0.620253164556962
20 0.680379746835443
30 0.661392405063291
40 0.639240506329114
50 0.563291139240506
60 0.579113924050633
70 0.528481012658228
80 0.531645569620253
90 0.572784810126582
100 0.572784810126582
110 0.59493670886076
120 0.572784810126582
130 0.509493670886076
140 0.401898734177215
150 0.272151898734177
160 0.237341772151899
170 0.215189873417722
180 0.218354430379747
190 0.237341772151899
200 0.237341772151899
210 0.310126582278481
220 0.367088607594937
230 0.474683544303797
240 0.525316455696203
250 0.556962025316456
260 0.566455696202532
270 0.496835443037975
280 0.503164556962025
290 0.503164556962025
300 0.560126582278481
310 0.620253164556962
320 0.636075949367089
330 0.626582278481013
340 0.693037974683544
350 0.664556962025316
360 0.436708860759494
}; \label{oscfar}
\addplot [line width=1pt, peru22113282]
table {%
0 0.35126582278481
10 0.607594936708861
20 0.645569620253165
30 0.661392405063291
40 0.572784810126582
50 0.642405063291139
60 0.620253164556962
70 0.585443037974684
80 0.515822784810127
90 0.69620253164557
100 0.585443037974684
110 0.677215189873418
120 0.737341772151899
130 0.598101265822785
140 0.465189873417722
150 0.471518987341772
160 0.506329113924051
170 0.5
180 0.395569620253165
190 0.512658227848101
200 0.506329113924051
210 0.487341772151899
220 0.436708860759494
230 0.60126582278481
240 0.60126582278481
250 0.65506329113924
260 0.651898734177215
270 0.64873417721519
280 0.572784810126582
290 0.632911392405063
300 0.613924050632911
310 0.575949367088608
320 0.604430379746835
330 0.585443037974684
340 0.598101265822785
350 0.591772151898734
360 0.35126582278481
};\label{deepest}
\addplot [line width=1pt, mediumseagreen85168104]
table {%
0 0
10 0.585443037974684
20 0.740506329113924
30 0.734177215189873
40 0.813291139240506
50 0.75
60 0.775316455696203
70 0.724683544303798
80 0.69620253164557
90 0.740506329113924
100 0.75
110 0.819620253164557
130 0.316455696202532
140 0.0537974683544304
150 0.0253164556962026
160 0.0158227848101266
170 0.0569620253164557
180 0
190 0
200 0.0664556962025317
210 0.0981012658227848
220 0.120253164556962
230 0.319620253164557
240 0.607594936708861
250 0.699367088607595
260 0.686708860759494
270 0.642405063291139
280 0.632911392405063
290 0.639240506329114
300 0.556962025316456
310 0.645569620253165
320 0.708860759493671
330 0.670886075949367
340 0.715189873417722
350 0.522151898734177
360 0
};\label{pymax}
\addplot [dashed, line width=1pt, steelblue76114176]
table {%
0 0.256329113924051
10 0.405063291139241
20 0.433544303797468
30 0.417721518987342
40 0.414556962025316
50 0.370253164556962
60 0.39873417721519
70 0.392405063291139
80 0.395569620253165
90 0.420886075949367
100 0.493670886075949
110 0.477848101265823
120 0.471518987341772
130 0.392405063291139
140 0.287974683544304
150 0.208860759493671
160 0.158227848101266
170 0.189873417721519
180 0.193037974683544
190 0.215189873417722
200 0.164556962025316
210 0.243670886075949
220 0.294303797468354
230 0.386075949367089
240 0.443037974683544
250 0.477848101265823
260 0.420886075949367
270 0.376582278481013
280 0.405063291139241
290 0.392405063291139
300 0.436708860759494
310 0.427215189873418
320 0.367088607594937
330 0.392405063291139
340 0.401898734177215
350 0.376582278481013
360 0.256329113924051
};\label{oscfar_small}
\addplot [dashed, line width=1pt, mediumseagreen85168104]
table {%
0 0
10 0.363924050632911
20 0.544303797468354
30 0.509493670886076
40 0.556962025316456
50 0.55379746835443
60 0.579113924050633
70 0.582278481012658
80 0.610759493670886
90 0.667721518987342
100 0.686708860759494
110 0.727848101265823
130 0.275316455696203
140 0.0537974683544304
150 0.0253164556962026
160 0.0158227848101266
170 0.0537974683544304
180 0
190 0
200 0.0569620253164557
210 0.0854430379746836
220 0.107594936708861
230 0.30379746835443
240 0.550632911392405
250 0.629746835443038
260 0.613924050632911
270 0.582278481012658
280 0.518987341772152
290 0.512658227848101
300 0.39873417721519
310 0.433544303797468
320 0.45253164556962
330 0.44620253164557
340 0.462025316455696
350 0.329113924050633
360 0
};\label{pymax_small}
\addplot [dashed, line width=1pt, peru22113282]
table {%
0 0.139240506329114
10 0.310126582278481
20 0.344936708860759
30 0.379746835443038
40 0.335443037974684
50 0.322784810126582
60 0.379746835443038
70 0.360759493670886
80 0.30379746835443
90 0.474683544303797
100 0.414556962025316
110 0.458860759493671
120 0.487341772151899
130 0.443037974683544
140 0.332278481012658
150 0.379746835443038
160 0.310126582278481
170 0.240506329113924
180 0.110759493670886
190 0.243670886075949
200 0.306962025316456
210 0.405063291139241
220 0.35126582278481
230 0.443037974683544
240 0.455696202531646
250 0.487341772151899
260 0.474683544303797
270 0.462025316455696
280 0.39873417721519
290 0.401898734177215
300 0.389240506329114
310 0.319620253164557
320 0.287974683544304
330 0.332278481012658
340 0.329113924050633
350 0.29746835443038
360 0.139240506329114
};\label{deepest_small}
\end{axis}

\matrix[
    matrix of nodes,
    anchor=north west,
    draw,
    inner sep=0.2em,
    draw=lightgray204,
    fill=white,
    fill opacity=0.8, draw opacity=1, text opacity=1
]
  at([xshift=-2.5cm, yshift=-0.1cm]current axis.north){
    \ref{oscfar}& OS-CFAR&
    \ref{los}& LoS Gain\\
    \ref{pymax}& RIMAX&&\\
    \ref{deepest}& DeepEst&&\\};

\end{tikzpicture}
    \caption{\textit{Empirical Target Detection Probability Compared to the \gls{los} Strength---While the solid lines depict this metric for an identification boundary of $\epsilon = 0.5$, the dashed lines represent $\epsilon = 0.25$.}}
    \vspace{-1em}
    \label{fig:4/detection_prob}
\end{figure}

\subsection{Comparison for One Bistatic Measurement Angle}
\label{4-comparison/one_angle}

We first compare the estimation results of RIMAX, DeepEst, and \gls{oscfar} for a bistatic observation angle of~\qty{20}{\degree}.
As this angle provides a bistatic scenario, the \gls{los} is at least~\qty{30}{\decibel} weaker compared to backward and forward scattering, as depicted by the red line in~\Cref{fig:4/detection_prob}.
This \gls{los} degradation arises from the horn antennas used in the measurement setup, which yield strong main and back lobes.
In a bistatic scenario, the \gls{los} of the \gls{tx} antenna is positioned outside the main and back lobes of the \gls{rx} antenna.
Consequently, the \gls{los} contribution to the data reduces, allowing all three algorithms to achieve a detection probability of at least~\qty{60}{\percent} for the two spheres.

\begin{figure*}[t]
    \centering
    \begin{subfigure}[t]{0.4\textwidth}
        \centering
        \input{figures/delay_doppler_plot/delay_py}
        \caption{\textit{Delay RIMAX}}
        \label{fig:del2}
    \end{subfigure}
    \hfill
    \begin{subfigure}[t]{0.28\textwidth}
        \centering
        \input{figures/delay_doppler_plot/delay_deep}
        \label{fig:del3}
        \caption{\textit{Delay DeepEst}}
    \end{subfigure}
    \hfill
    \begin{subfigure}[t]{0.28\textwidth}
        \centering
        \input{figures/delay_doppler_plot/delay_per}
        \caption{\textit{Delay \gls{oscfar}}}
        \label{fig:del1}
    \end{subfigure}
    \\[1em]
    \begin{subfigure}[t]{0.4\textwidth}
        \centering
        \input{figures/delay_doppler_plot/doppler_py}
        \caption{\textit{Doppler RIMAX}}
        \label{fig:dop2}
    \end{subfigure}
    \hfill
    \begin{subfigure}[t]{0.28\textwidth}
        \centering
        \input{figures/delay_doppler_plot/doppler_deep}
        \caption{\textit{Doppler DeepEst}}
        \label{fig:dop3}
    \end{subfigure}
    \hfill
    \begin{subfigure}[t]{0.28\textwidth}
        \centering
        \input{figures/delay_doppler_plot/doppler_per}
        \caption{\textit{Doppler \gls{oscfar}}}
        \label{fig:dop1}
    \end{subfigure}
    \caption{\textit{Delay-Doppler Estimation Results for a Bistatic Observation Angle of $\delta = \qty{20}{\degree}$---All plots include the results for one full rotation of the target emulator. DeepEst estimates a large number of static paths  due to the missing background subtraction.}}
    \vspace{-1.5em}
    \label{fig:4/delay_doppler}
\end{figure*}

\Cref{fig:4/delay_doppler} depicts the temporal progression of the delay (top row) and Doppler (bottom row) estimation results for the three algorithms.
Due to applied background subtraction, the estimation results of \gls{oscfar} and RIMAX do not incorporate any static paths.
Consequently, \Cref{fig:dop2} and \Cref{fig:dop1} only depict a negligible number of detections that do not stem from the two spheres.
DeepEst, on the other hand, does not include background subtraction and, therefore, detects mostly static paths.
This accumulation around a Doppler-shift of~\qty{0}{\hertz} causes delay estimations, where most detections do not align with the delay-Doppler ground truth of the spheres.
In fact, the results of DeepEst show reflections from the measurement chamber wall and the motor housing, for example at a delay of approximately~\qty{35}{\nano\second}~\cite{Schwind.2020}.

As illustrated in Figures~\mbox{\ref{fig:dop2}-\ref{fig:dop1}}, the Doppler estimations of each algorithm reveal the contribution of a linear frequency path emerging between~\qty{0.2}{\second} and~\qty{0.4}{\second}, and between~\qty{0.7}{\second} and~\qty{0.9}{\second}.
This path originates from a reflection of the transmitted signal at the target emulator beam.
As this beam rotates, the specular reflection center moves across its surface, resulting in the linear Doppler progression.

\subsection{Comparison for All Bistatic Measurement Angles}
\label{4-comparison/performance_comparison}

As the bistatic measurement angle affects both the \gls{los} strength and the absolute Doppler-shifts of the spheres, separating the spheres from the static paths becomes more challenging as this angle approaches~\qty{180}{\degree}.
\Cref{fig:4/detection_prob} depicts the dependency of the detection probability on the bistatic measurement angle for \gls{oscfar} (blue), RIMAX (green), and DeepEst (orange).
The red line visualizes the strength of the \gls{los}, which appears inversely proportional to the empirical target detection probability.
In essence, the figure reveals that all three algorithms share a similar dependency on the bistatic observation angle.
Backward and forward scattering conditions pose a problem for sphere detections, yielding low probabilities down to~\qty{0}{\percent} for RIMAX.
Conversely, bistatic scattering scenarios provide a higher number of true detections, with RIMAX being capable of estimating the two targets in more than~\qty{80}{\percent} of cases.
The poor detection rates of RIMAX in backward and forward scattering are attributable to errors in the measurement hardware, which predominate in the delay-Doppler spectrum for strong \gls{los} scenarios.

In addition, the availability of ground truth values enables the calculation of \glspl{rmse} utilizing~\eqref{eq:5/rmse}, with the caveat that only successful sphere detections are included in this calculation.
\Cref{fig:4/rmse} depicts these delay and Doppler \glspl{rmse}.
It can be observed that the \glspl{rmse} for the three algorithms under consideration are comparable in both magnitude and progression across the full sweep of the bistatic observation angle.
This finding aligns with the results presented in \Cref{fig:4/delay_doppler}, which indicate that some estimates closely follow the progression of the delay-Doppler ground truth for most time points.
However, Figure~\ref{fig:4/rmse} exhibits anomalies when $\delta$ approaches~\qty{180}{\degree}.
While the \glspl{rmse} of the proposed algorithm and RIMAX continue to decrease, those of DeepEst exhibit a maximum at this angle.
This effect is caused by the lack of background subtraction in DeepEst.
Consequently, DeepEst labels a significant number of static paths as sphere reflections.
These falsely classified paths are farther from the ground truth values, resulting in increased \glspl{rmse}.

\begin{figure}[ht]
    \centering
    \begin{tikzpicture}
\definecolor{orchid218139195}{RGB}{218,139,195}
\definecolor{darkslategrey38}{RGB}{38,38,38}
\definecolor{grey14712096}{RGB}{147,120,96}
\definecolor{indianred1967882}{RGB}{196,78,82}
\definecolor{lavender234234242}{RGB}{234,234,242}
\definecolor{lightgrey204}{RGB}{204,204,204}
\definecolor{lightslategrey129114179}{RGB}{129,114,179}
\definecolor{mediumseagreen85168104}{RGB}{85,168,104}
\definecolor{peru22113282}{RGB}{221,132,82}
\definecolor{steelblue76114176}{RGB}{76,114,176}

\definecolor{darkgray176}{RGB}{176,176,176}
\definecolor{lightgray204}{RGB}{204,204,204}

\begin{axis}[
width=0.9\columnwidth,
height=0.7\columnwidth,
legend cell align={left},
legend style={fill opacity=0.8, draw opacity=1, text opacity=1, draw=lightgray204},
tick align=outside,
tick pos=left,
x grid style={darkgray176},
xlabel={Bistatic Measurement Angle $\delta$ / \si{\degree}},
xmajorgrids,
xmin=0, xmax=360,
xtick style={color=black},
y grid style={darkgray176},
ylabel={RMSE Delay / \si{\nano\second}},
ymajorgrids,
ymin=0, ymax=2.4e-09,
ytick style={color=black},
ytick={0e-10,6e-10,12e-10,1.8e-09},
yticklabels={0,0.6,1.2,1.8},
scaled y ticks=false,
axis y line=left
]
\addplot [line width=1pt, steelblue76114176]
table {%
0 8.70279971508685e-10
10 9.44144752296989e-10
20 1.01791407021238e-09
30 9.70871355531807e-10
40 9.6006811000246e-10
50 9.93426562895354e-10
60 9.33821656301879e-10
70 7.96459779510642e-10
80 8.0152857735375e-10
90 8.41447115536389e-10
100 8.06428121153989e-10
110 8.31380236057134e-10
120 8.26767989017438e-10
130 8.42226852013951e-10
140 9.97835801196434e-10
150 7.93993856093292e-10
160 7.48491355232878e-10
170 5.18305101209929e-10
180 3.84915959369312e-10
190 4.9923194237719e-10
200 7.59920470469108e-10
210 8.01319344220074e-10
220 8.87445361275803e-10
230 8.51231626490294e-10
240 7.61607573166475e-10
250 7.79698266120421e-10
260 9.14809651890751e-10
270 7.83162491521166e-10
280 8.60047780609285e-10
290 9.02765708773565e-10
300 7.83804256905557e-10
310 8.97739706000751e-10
320 9.36316385110352e-10
330 9.08348544865779e-10
340 9.7626762219768e-10
350 1.06376570614199e-09
360 8.70279971508685e-10
};\label{per_del}
\addplot [line width=1pt, peru22113282]
table {%
0 1.48329403400048e-09
10 1.21277840696123e-09
20 1.20520148523231e-09
30 1.13268763723102e-09
40 1.01641611635908e-09
50 1.22548297332264e-09
60 1.12805535659923e-09
70 1.2102077191223e-09
80 1.10931017604714e-09
90 1.00955220359381e-09
100 1.0063106811868e-09
110 1.06565694429189e-09
120 1.16638340962053e-09
130 8.17700742439289e-10
140 1.09682923336095e-09
150 8.597731082579e-10
160 7.11615850904367e-10
170 6.76982603425006e-10
180 1.04124440488062e-09
190 6.8894334537541e-10
200 6.75198975294214e-10
210 7.15263050307914e-10
220 9.48207103505182e-10
230 9.21029335719219e-10
240 9.81438847351762e-10
250 1.07904121771521e-09
260 1.07007885256365e-09
270 1.12164491206776e-09
280 9.47173474333602e-10
290 1.10392594395243e-09
300 1.06407234604772e-09
310 1.14046723573404e-09
320 1.25636635728034e-09
330 1.08165710981554e-09
340 1.18473758522467e-09
350 1.15782136201611e-09
360 1.48329403400048e-09
};\label{deep_del}
\addplot [line width=1pt, mediumseagreen85168104]
table {%
0 nan
10 1.02496121343616e-09
20 9.6250265948764e-10
30 1.0291135757286e-09
40 9.68650887222949e-10
50 9.13960218176178e-10
60 9.23341715185007e-10
70 8.0347365161584e-10
80 8.25783045327508e-10
90 7.32624198505337e-10
100 8.06962168321297e-10
110 8.12237037851066e-10
130 7.66912077548568e-10
140 6.56971279714198e-10
150 7.15751370967284e-10
160 7.62260046003276e-10
170 5.30996360164842e-10
180 nan
190 nan
200 3.88266922172031e-10
210 7.02331526091043e-10
220 7.89301541778903e-10
230 6.42938240320535e-10
240 7.40162533366044e-10
250 8.47247697111189e-10
260 9.15381626373223e-10
270 7.05217055667495e-10
280 8.55722003117161e-10
290 7.90171764407739e-10
300 8.98134172183834e-10
310 1.05290741044403e-09
320 9.97336941476541e-10
330 8.97058584818763e-10
340 9.70991279931433e-10
350 8.71258278735285e-10
360 nan
};\label{py_del}
\end{axis}

\begin{axis}[
width=0.9\columnwidth,
height=0.7\columnwidth,
legend cell align={left},
legend style={fill opacity=0.8, draw opacity=1, text opacity=1, draw=lightgray204},
tick align=outside,
tick pos=right,
axis y line = right,
axis x line = none,
x grid style={darkgray176},
xlabel={TX-RX angle [°]},
xmajorgrids,
xmin=0, xmax=360,
xtick style={color=black},
y grid style={darkgray176},
ylabel={RMSE Doppler / \si{\hertz}},
ymajorgrids,
ymin=0, ymax=80,
ytick style={color=black},
ytick={0, 20, 40, 60},
yticklabels={0, 20, 40, 60},
]
\addplot [dashed, line width=1pt, steelblue76114176]
table {%
0 40.5918187597687
10 36.325838858407
20 35.0759962367265
30 33.4245947246264
40 32.5062408763405
50 33.378453643348
60 32.4968923250797
70 31.0608052052612
80 32.4121360373695
90 31.5745026405984
100 26.3195497945492
110 26.8028254903292
120 26.8485714118877
130 29.7736415707106
140 29.4342944517874
150 28.7898048411698
160 36.2161450282191
170 22.8489159022749
180 22.875282164365
190 23.120677383249
200 32.7751968120768
210 28.0041419507796
220 28.2577714762893
230 28.2857779401034
240 25.9485940012272
250 27.2208412461159
260 30.6164126811662
270 30.7312356869009
280 30.7040195586795
290 30.078191761555
300 30.7471579934772
310 34.0811950319032
320 34.9616096485393
330 35.6265669795142
340 36.8759487674996
350 37.633068883659
360 40.5918187597687
};\label{per_dop}
\addplot [dashed, line width=1pt, peru22113282]
table {%
0 37.6109873120406
10 39.4729565027883
20 37.704622710779
30 36.3753582215473
40 35.8120020356375
50 36.6514857378849
60 35.1704781579124
70 33.3654149861295
80 36.516178027219
90 34.100526180587
100 32.1379425387996
110 31.0726414567613
120 30.0727200209609
130 30.5763145656811
140 30.4427183968427
150 29.8836778666094
160 35.2255288913145
170 43.7582709248508
180 48.8848936509064
190 43.7232213180654
200 35.5138320862518
210 29.9057022449071
220 25.8891909048141
230 29.4377379224354
240 27.0359183508744
250 26.8686510333081
260 28.7702243902285
270 29.715184703191
280 33.4527629184865
290 32.8094424658073
300 35.5049845210882
310 38.1476345436238
320 37.7268432513121
330 37.6770983222234
340 37.638559036954
350 39.3971365929292
360 37.6109873120406
};\label{deep_dop}
\addplot [dashed, line width=1pt, mediumseagreen85168104]
table {%
0 nan
10 37.6737549434871
20 32.9690228679892
30 32.6472232774535
40 32.0099694009116
50 30.5212211371006
60 29.488790990182
70 26.3783883938662
80 24.5178228061284
90 24.3371458945317
100 21.1555163361905
110 21.6474084244724
130 26.9828655003617
140 13.4974989153463
150 13.3472866842761
160 20.7880414430905
170 26.2826484722317
180 nan
190 nan
200 26.0657448343385
210 26.5243403135279
220 22.9237629699249
230 20.3563231442668
240 21.7127340583779
250 21.8478716720445
260 23.1956989779435
270 24.4141359779222
280 28.1890352026665
290 27.3623440392949
300 31.5377612801544
310 33.4836634916725
320 33.2872747391015
330 35.2028523705937
340 35.484675170104
350 37.6954788215495
360 nan
};\label{py_dop}

\end{axis}

\matrix[
    matrix of nodes,
    anchor=north west,
    draw,
    inner sep=0.2em,
    draw=lightgray204,
    fill=white,
    fill opacity=0.8, draw opacity=1, text opacity=1,
    nodes={anchor=west}
]
  at([xshift=-2.3cm, yshift=-0.1cm]current axis.north){
    \ref{oscfar_solid}& OS-CFAR & \ref{oscfar_dashed}& Delay\\
    \ref{pymax_solid}& RIMAX & \ref{pymax_dashed}& Doppler\\
    \ref{deepest_solid}& DeepEst &&\\};

\end{tikzpicture}
    \vspace{-1em}
    \caption{\textit{Delay and Doppler \glspl{rmse}---\gls{oscfar} (blue), RIMAX (green), and DeepEst (orange) exhibit similar \glspl{rmse} both in magnitude and progression. Due to an incorrect classification of static paths, the latter algorithm yields increased errors in forward scattering scenarios.}}
    \vspace{-1em}
    \label{fig:4/rmse}
\end{figure}

Finally, we perform a runtime analysis of the three algorithms, by comparing their runtimes in terms of the mean execution time per radar frame, which is of size $1024\times100$.
Given that the numerical estimation of the runtime is subject to the specifications of the utilized hardware, we provide these results relative to the mean execution time of RIMAX.
In essence, \gls{oscfar} and DeepEst process a radar frame \qty{95.30}{\percent} and \qty{92.46}{\percent} faster than RIMAX, respectively.
\section{Conclusion}
\label{5-conclusion}
Within our work, we compare \gls{ml}-, \gls{cnn}-, and \gls{oscfar}-based delay-Doppler estimation algorithms based on the empirical target detection probability, \glspl{rmse}, and relative computation time.
We show that the empirical target detection probabilities demonstrate similar responses to relative changes in the magnitude of the \gls{los}.
In essence, an increase in the \gls{los} yields a reduction in the empirical target detection probability and vise versa.
In addition, we demonstrate that the \gls{oscfar}-based estimator offers higher computational efficiency than \gls{ml}- and \gls{cnn}-based approaches.
However, owing to their high-resolution capabilities, the latter two algorithms provide more accurate estimates when the targets are masked by stronger static paths.

To validate the results of our comparison, the presented algorithms should be applied to measurement data obtained outside a controlled environment.
In such scenarios, which feature a higher number of targets as well as increased clutter and noise levels, \gls{ml}- and \gls{cnn}-based algorithms are expected to perform better due to their high-resolution capabilities and iterative cancelation procedures.
\printbibliography

\end{document}